\documentstyle [art11,epsf]{article}
\title	{$SU(N_c)$ gauge theories for all $N_c$ \\
in 3 and 4 dimensions.}
\date{}
\author	{
\\
M. Teper 
\\
{\small\sl Theoretical Physics, University of Oxford,
	1 Keble Road, Oxford, OX1 3NP, U.K.}}
\vspace{3ex}
\begin	{document}
\maketitle
\vspace{3ex}
\vspace{3ex}
\begin	{abstract}
\noindent
We compare the mass spectra and string tensions
of $SU(2)$, $SU(3)$ and $SU(4)$ gauge theories in 
2+1 dimensions. We find that the ratios of masses are, to a
first approximation, independent of $N_c$ and that the 
remaining dependence can be accurately reproduced by a simple 
$O(1/N^2_c)$ correction. This provides us with a prediction 
of these mass ratios for all $SU(N_c)$ theories in 2+1 
dimensions and demonstrates that these theories
are `close' to $N_c=\infty$ for $N_c\geq 2$. We also
find that, when expressed in units of the dynamical length scale
of the theory, the dimensionful coupling $g^2$ is proportional 
to $1/N_c$ at large $N_c$. We confirm that these
theories are indeed confining in the limit $N_c \to \infty$.
We describe preliminary calculations in 3+1 dimensions which 
indicate that the same will be true there. 
\end	{abstract}

\vfill
Oxford Preprint Number: {\em OUTP--97--01P}
\hfill
hep-lat/9701003

\newpage
\section	{Introduction.}
\label		{intro}

The proposal to consider $SU(N_c)$ gauge  
theories as perturbations in powers of $1/N_c$ 
around $N_c=\infty$ is an old one
\cite{tHo}.
If one assumes 
confinement for all $N_c$, then the phenomenology
of the $SU(\infty)$ quark-gluon theory is strikingly 
similar to that of (the non-baryonic sector of) QCD
\cite{tHo,Wit}.
This makes it conceivable that the physically
interesting $SU(3)$ theory could be largely
understood by solving the much simpler $SU(\infty)$ theory
\cite{Cole}.
The fact that the lattice $SU(\infty)$ theory
can be re-expressed as a single plaquette
theory
\cite{EK},
has provided the basis of a number of interesting
computational explorations (for a review see
\cite{Das}).
Unfortunately this latter scheme makes no statement
about the size of even the leading corrections to the
$N_c=\infty$ limit, and so gives us no clue as to
how close $N_c=3$ is to $N_c=\infty$.

In this paper we calculate the properties of $SU(N_c)$ gauge 
theories for several values of $N_c$ and explicitly determine 
how the physics varies as $N_c$ increases. 
We note that there are good reasons for believing that the 
inclusion of quarks would not alter any of our conclusions
(except in some obvious ways). We 
consider both 2+1 and 3+1 dimensions. The former calculations are 
much the more precise and it is there that we will be able to make 
some firm statements. In $D=3+1$ our conclusions will be 
similar but more tentative.

While one might naively expect that the $D=2+1$ and $D=3+1$ 
gauge theories would be so different as to make a unified
treatment misleading, this is not in fact so. Theoretically
the $D=2+1$ theory shares with its $D=3+1$ homologues
three central properties. Firstly, at short distances,
the dimensionless coupling becomes weak in both cases.
In the $D=2+1$ theory the coupling, $g^2$ has dimensions
of mass so that the effective dimensionless expansion
parameter on a scale $a$ will be $ag^2$ which vanishes
linearly with distance (the theory is super-renormalisable).
In the $D=3+1$ theory the coupling vanishes logarithmically
with distance (asymptotic freedom). Secondly, at large
distances both theories appear to be confining, with a
non-perturbative linear potential between fundamental sources. 
Thirdly it is the value of the coupling that sets the
overall mass scale in both cases. In $D=2+1$ this arises directly
because $g^2$ has dimensions of mass. In $D=3+1$ it does so 
through the phenomenon of dimensional transmutation: the classical 
scale invariance is anomalous, the coupling runs and this introduces 
a mass scale through the rate at which it runs
(i.e. the $\Lambda_{\overline{MS}}$ parameter). In addition to these 
general theoretical similarities, the calculated spectra also
show some striking similarities. All this motivates us
to believe that a unified treatment makes sense.

The $D=2+1$ analysis is based on our calculations over the last 
few years of the properties of $SU(N_c)$ gauge theories with
$N_c=2,3$ and 4. In $D=3+1$ what we have done is to perform
some $SU(4)$ calculations to supplement what is known about
$SU(2)$ and $SU(3)$. Our strategy is the very simple one of
directly calculating the mass spectra of these theories and
seeing whether they are approximately independent of $N_c$.
The calculations are performed through the Monte Carlo
simulation of the corresponding lattice theories. In the $D=2+1$ 
case the calculations are very accurate and we are able to 
extrapolate our mass ratios to the continuum limit prior to
the comparison. In the 3+1 dimensional case our $SU(4)$
calculations are not good enough for that, and our
comparisons with $SU(2)$ and $SU(3)$ are correspondingly
less precise.

This study was originally motivated by the observation that
the $C=+$ sector of the light mass spectrum turned 
out to be quite similar in the $D=2+1$ $SU(2)$ 
\cite{MT2G}
and $SU(3)$
\cite{MT3G}
theories. (This also appears to be the case in $D=3+1$, 
although there the comparison is weakened by 
the much larger errors.) If the reason for this is
that both are close to the $N_c=\infty$ limit, then
this provides an economical understanding of the spectra
of $SU(N_c)$ gauge theories for all $N_c$, i.e.
there is a common spectrum with small corrections.

A second reason for studying $N_c \to \infty$  is
that models and theoretical approaches are usually simpler
in that limit. For example, the flux tube model 
of glueballs
\cite{patisg,tepmor}
would naively appear to be identical for $N_c > 2$.
However, because the model does not incorporate the
effects of glueball decay, it should in fact be tested
against the $N_c\to\infty$ spectrum since it is only in that 
limit that there are no decays. A second example 
is provided by the recent progress in calculating the
large $N_c$ mass spectrum using light-front quantisation
techniques
\cite{Dalley}.

The calculations in this paper have all been performed with
the standard plaquette action using standard Monte Carlo
techniques. The lattice spacing, $a$, is varied by 
changing the dimensionless bare inverse coupling, $\beta$,
which appears in the lattice action. In the pure gauge
theory the states are necessarily composed entirely of
gluons and we shall therefore refer to them as `glueballs'.
Some of the $SU(2)$ results 
have been published 
\cite{MT2G,MT2K}
and a paper on the other work is in preparation. A brief
summary of the work in this paper has been presented
elsewhere
\cite{MTLAT96}.

\section	{2+1 dimensions.}

\label	{3dim}

We begin with the string tension, $\sigma$, since it turns out 
to be the most accurate physical quantity in our calculations. 
We use smeared Polyakov loops 
\cite{CMT},
to obtain $a^2\sigma$ for several values of the lattice spacing $a$.
We then extrapolate the lattice results, using the asymptotic
relation $\beta = 4/ag^2$, to obtain 
the continuum string tension in units of $g^2$:
$$ {{\surd \sigma} \over g^2} = 
\lim_{\beta\to\infty} {\beta \over 4} a\surd\sigma
\eqno(1) $$
The results for $SU(2)$
\cite{MT2K},
$SU(3)$ and $SU(4)$
\cite{MT3G},
in $D=2+1$ are shown in Table~\ref{n_string}.
We immediately see that there is an
approximate linear rise with $N_c$ and we find that we
can obtain a good fit with
$$ {{\surd \sigma} \over g^2} = 0.1974(12) N_c 
-{0.120(8) \over N_c}. \eqno(2) $$
We obtain a similar behaviour with the light
glueball masses (see below).

Some observations.

$\bullet$ For large $N_c$, eqn(2) tells us that 
$\surd\sigma \propto g^2 N_c$. That is to say, 
the overall mass scale of the theory, call
it $\mu$, is proportional to $g^2 N_c$. In other words, 
in units of the mass scale of the theory
$$ g^2 \propto {\mu \over N_c}. \eqno(3)$$
While this coincides with the usual expectation based on an 
analysis of Feynman diagrams, we note that here the
argument is fully non-perturbative.

$\bullet$ The string tension is non-zero for all $N_c$ 
and, in particular, for $N_c \to \infty$ (when expressed
in units of $g^2 N_c$ or the lightest glueball masses - see
below). This confirms the basic assumption that needs to
be made in 4 dimensions
in order to extract the usual phenomenology of the 
large-$N_c$ theory
\cite{Cole,Wit}.

$\bullet$ In the pure gauge sector one expects (again
from an analysis of Feynman diagrams) 
\cite{Cole}
that the first correction to the large-$N_c$ limit will be
$O(1/N^2_c)$ relative to the leading term. The fit in
eqn(2) is indeed of this form. We note that if we try a fit 
with a $O(1/N_c)$ correction instead (which would be 
appropriate if we had quarks) then we obtain an unacceptably 
poor $\chi^2$ (corresponding to a confidence level of only 
$\sim 2\%$ in contrast to the $\sim 45\%$ we obtain for the
quadratic correction). We may regard this as providing some 
non-perturbative support for the diagram-based expectation.

$\bullet$ The coefficient of the correction term is 
comparable to that of the leading term, suggesting an
expansion in powers of $1/N_c$ that is rapidly convergent.
Indeed one has to go to $N_c=1$ before the correction term 
becomes comparable to the leading term. While the $SU(1)$
theory is completely trivial, we note that the $U(1)$
theory has a zero string tension (in the sense that
$\surd\sigma/g^2 = 0$ in the continuum limit).

In addition to the string tension we have calculated 
part of the mass spectrum. In particular we have
calculated the masses of the lightest particles
with $J^{PC}$ quantum numbers $J=0,1,2$, $P=\pm$
and $C=\pm$. In some cases we have calculated some
of the excited states for given $J^{PC}$. Note
that since we are in $D=2+1$, states of opposite
parity are degenerate as long as $J \not = 0$.
This degeneracy is broken by lattice spacing and
finite volume corrections. We will present our
results separately for the $P=+$ and $P=-$ states
so as to provide an explicit check on the presence of any
such unwanted corrections.

\begin	{figure}[p]
\begin	{center}
\leavevmode
\input	{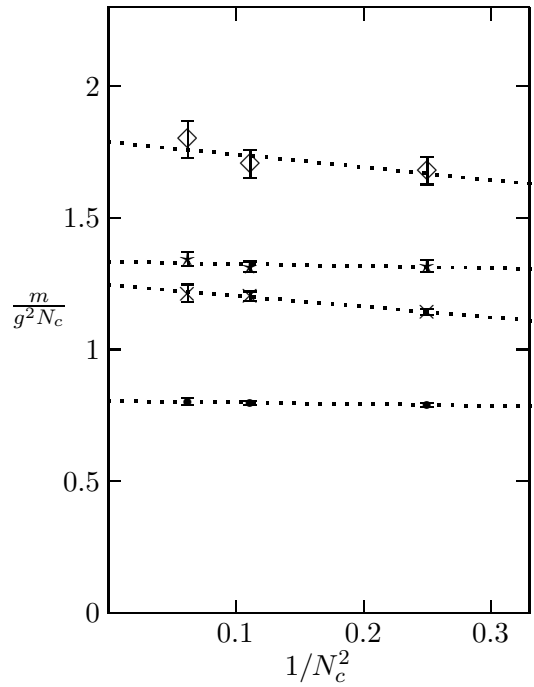}
\end	{center}
\caption{Some continuum glueball masses, in $D=3$, for 2,3,4
colours: $0^{++}$($\bullet$), $0^{++*}$($\times$), $2^{++}$($\star$), 
$0^{-+}$($\diamond$) and linear fits. }

\label	{fig_plot_glue3g}
\end 	{figure}

We begin with the $C=+$ spectrum since the $SU(2)$ spectrum
does not contain $C=-$ states. In this case we have
masses for three values of $N_c$, and so can check
how good is a fit of the kind in eqn(2). In
Fig.~\ref{fig_plot_glue3g}.
we plot the ratio $m_G/g^2N_c$ against $1/N^2_c$ for a selection
of the lightest states, $G$. On this plot a fit of the form
in eqn(2) will be a straight line and we show the best such fits.
 As we can see, the data is
consistent with such a $1/N^2_c$ correction being
dominant for $N_c \geq 2$. However what is 
really striking is the lack of $any$ apparent $N_c$
dependence for the lightest $0^{++}$ and $2^{++}$
states. 

In Table~\ref{ng_plus} we present the results of fitting the
$C=+$ states to the form
$$ 
{{m_G}\over{g^2 N_c}}
= R_{\infty} + {R_{slope} \over N^2_c}. 
\eqno(4)$$
where $R_{\infty}={{m_G}\over{g^2 N_c}}{\Bigl/}_{N_c=\infty}$.
(Note that the errors on the slope and intercept are highly
correlated.) For each state we show the confidence level of the
fit. These are  acceptable 
suggesting once again that for $N_c \geq 2$ a moderately 
sized correction of the form $\sim 1/N^2_c$ is all that
is needed. Note that since the variation with $N_c$ is small,
the exact form of the correction used will not have
a large impact on the extrapolation to $N_c=\infty$
(except in estimating the errors).

These calculations confirm our earlier claim that the physical 
mass scale at large $N_c$ is $g^2N_c$. So if we consider ratios
of $m_G$ to $\surd\sigma$ (as was explicitly
done in 
\cite{MTLAT96})
we will find that they have finite non-zero limits as
$N_c\to\infty$ : that is to say, the large-$N_c$ theory 
possesses linear confinement.

For the $C=-$ states we only have masses for 2 values of
$N_c$ and we cannot therefore check whether a fit
of the form in eqn(4) is statistically favoured or not. 
However given that such a fit has proved accurate for 
the $C=+$ masses and for the string tension down to
$N_c=2$ it seems entirely resonable to assume that
it will be appropriate for $N_c \geq 3$ for the
$C=-$ masses. Assuming this we obtain the results
shown in Table~\ref{ng_minus} for the $N_c = \infty$ limit
and for the coefficient of the first correction.
The `lever arm' on this extrapolation is, of course, shorter 
than for the $C=+$ states and that leads to correspondingly 
larger errors.

The results in the Tables provide us not only with values
for the various mass ratios in the limit $N_c\to\infty$
but also, when inserted into eqn(2), predictions 
for $all$ values of $N_c$.
 
Finally we remark that we have also calculated the deconfining
temperature, $T_c$, for $SU(2)$
\cite{tcsu2}
and for $SU(3)$
\cite{MT3G}.
Extrapolating as in eqn(4), we find
$$ {T_c \over {g^2 N_c}} = 0.1745(52)
+{0.079(23) \over N^2_c}. \eqno(5) $$
Of course, extrapolating from $N_c=2,3$ is less reliable
than extrapolating from $N_c=3,4$.

\section	{3+1 dimensions.}

\label	{4dim}

Our knowledge of 4 dimensional gauge theories is much less 
precise. As far as continuum properties are concerned, 
quantities that are known with reasonable accuracy include the 
string tension, the lightest scalar and tensor glueballs, the
deconfining temperature and the topological susceptibility. 
As in 3 dimensions, the $SU(2)$ and $SU(3)$ values are 
within $\sim 20\%$ of each other, which encourages us
to investigate the $SU(4)$ theory so as to see whether
we are indeed `close' to $N_c=\infty$. Of course
$SU(4)$ calculations are much slower in $D=3+1$ and the
results we present here are of a preliminary nature.

We use the standard plaquette action, and so our
first potential hurdle is the presence of the well-known 
bulk transition that occurs as we increase the inverse 
bare coupling, $\beta \equiv 2N_c/g^2$, from strong 
towards weak coupling. To locate this transition we performed 
a scan on a $10^4$ lattice and found that it occurred
at $\beta = 10.4 \pm 0.1$. This corresponds to
a rather large value of the lattice spacing, $a$,
and so does not lie in the range of couplings within which 
we shall be working, i.e. $\beta=$10.7,10.9 and 11.1.

Our calculation consists of 4000,6000 and 3000 sweeps on 
$10^4$,$12^4$ and $16^4$ lattices at $\beta=$10.7,10.9 and 
11.1 respectively. Every fifth sweep we calculated correlations 
of (smeared) gluonic loops and from these we extracted
the string tension and the masses of the lightest 
$0^{++}$ and $2^{++}$ particles, using standard techniques
\cite{CMT}.
These are presented in Table~\ref{ngk_4d}. We also calculated
the topological susceptibility, $a^4\chi_t \equiv <Q^2>/L^4$,
where $L^4$ is the number of lattice sites and $Q$ is the
total topological charge. (Note that in $D=2+1$ there is no
such charge.) The charge $Q$ was obtained using a standard
cooling method, just as in $SU(3)$
\cite{Qsu3}.
The calculations were performed every 50 sweeps. Overall this
corresponds to rather small statistics and the errors are
therefore unlikely to be very reliable. 

We see from Table~\ref{ngk_4d} that the most accurate 
physical quantity in our calculations is the string tension, 
$\sigma$. Can we learn from it how $g^2$ varies with $N_c$, just 
as we did in $D=2+1$? We focus on a simple aspect of this question:
if we compare different $SU(N_c)$ theories at a value of
$a$ which is the same in physical units, i.e. for which
$a\surd\sigma$ is the same, does the bare coupling
vary as $1/N_c$, i.e. does $\beta \equiv 2N_c/g^2 \propto N^2_c$?
We perform this comparison for $\beta_4=10.9,11.1$. (For
convenience we shall label $\beta$ by the value
of $N_c$, i.e. we write it as $\beta_{N_c}$.)
To find the corresponding values of $\beta$ in
$SU(2)$ and $SU(3)$ we simply interpolate between the
values provided in (for example)
\cite{CMT}.
Doing so we find that the values of $\beta$ corresponding
to $\beta_4=10.9,11.1$ are  $\beta_3 \simeq 5.972(18),6.071(24)$ 
and $\beta_2 \simeq 2.442(9),2.485(11)$ respectively. If we
simply scale $\beta_4$ by $N^2_c$ then what we would have
expected to obtain is  $\beta_3 \simeq 6.131,6.244$ 
and $\beta_2 \simeq 2.725,2.775$ respectively. Superficially
the numbers look to be in the right ballpark, but
in fact the agreement is poor. For example 
$\beta_2=2.725$ and $\beta_2=2.442$ correspond to
values of $a\surd\sigma$ that differ by about a
factor of 3. 

This disagreement should not, however, be taken too
seriously, since it is well-known that the lattice
bare coupling is a very poor perturbative expansion parameter.
It is known that one can get a much better expansion
parameter if one uses instead the mean-field improved coupling,
$g^2_I$, obtained from $g^2$ by dividing
it by the average plaquette, $<{1\over N_c}TrU_p>$ 
\cite{MFI}.
Defining $\beta^I_{N_c} \equiv 2N_c/g^2_I(a)$ we find
that $\beta_4=10.9, 11.1$ correspond to $\beta^I_4=6.215, 6.474$
respectively. Scaling $\beta^I_4$ by $N^2_c$ we would
expect the equivalent $SU(3)$ and $SU(2)$ couplings to be
given by $\beta^I_3=3.496, 3.642$ and  $\beta^I_2=1.554, 1.619$.
What we actually find is that the equivalent couplings are
$\beta^I_3 \simeq 3.527(22), 3.649(28)$  and
$\beta^I_2=1.561(10), 1.613(12)$. The agreement is now
excellent. That is to say, if the $SU(N_c)$ mean-field 
improved bare-coupling is defined on a length scale that is
related to the physical length scale ($\surd\sigma$) by some
constant factor, then it varies as $g^2 \propto 1/N_c$.
This is, of course, the usual diagram-based expectation.

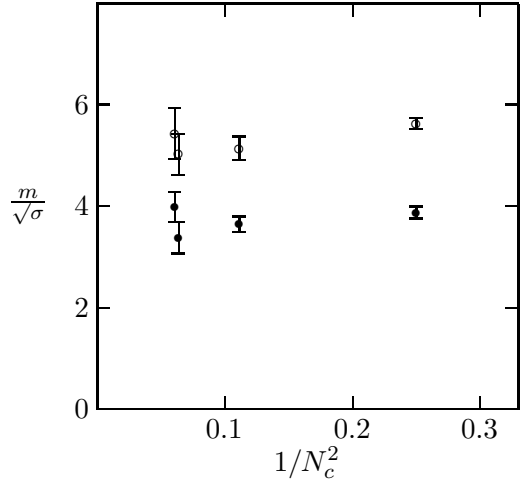
\begin	{figure}[p]
\begin	{center}
\leavevmode
\setlength{\unitlength}{0.240900pt}
\ifx\plotpoint\undefined\newsavebox{\plotpoint}\fi
\sbox{\plotpoint}{\rule[-0.200pt]{0.400pt}{0.400pt}}%
\begin{picture}(825,675)(0,0)
\font\gnuplot=cmr10 at 12pt
\gnuplot
\sbox{\plotpoint}{\rule[-0.200pt]{0.400pt}{0.400pt}}%
\put(120.0,31.0){\rule[-0.200pt]{4.818pt}{0.400pt}}
\put(108,31){\makebox(0,0)[r]{{$0$}}}
\put(761.0,31.0){\rule[-0.200pt]{4.818pt}{0.400pt}}
\put(120.0,190.0){\rule[-0.200pt]{4.818pt}{0.400pt}}
\put(108,190){\makebox(0,0)[r]{{$2$}}}
\put(761.0,190.0){\rule[-0.200pt]{4.818pt}{0.400pt}}
\put(120.0,350.0){\rule[-0.200pt]{4.818pt}{0.400pt}}
\put(108,350){\makebox(0,0)[r]{{$4$}}}
\put(761.0,350.0){\rule[-0.200pt]{4.818pt}{0.400pt}}
\put(120.0,509.0){\rule[-0.200pt]{4.818pt}{0.400pt}}
\put(108,509){\makebox(0,0)[r]{{$6$}}}
\put(761.0,509.0){\rule[-0.200pt]{4.818pt}{0.400pt}}
\put(320.0,31.0){\rule[-0.200pt]{0.400pt}{4.818pt}}
\put(320,19){\makebox(0,0){\shortstack{\\ \\ \\ {$0.1$}}}}
\put(320.0,648.0){\rule[-0.200pt]{0.400pt}{4.818pt}}
\put(521.0,31.0){\rule[-0.200pt]{0.400pt}{4.818pt}}
\put(521,19){\makebox(0,0){\shortstack{\\ \\ \\ {$0.2$}}}}
\put(521.0,648.0){\rule[-0.200pt]{0.400pt}{4.818pt}}
\put(721.0,31.0){\rule[-0.200pt]{0.400pt}{4.818pt}}
\put(721,19){\makebox(0,0){\shortstack{\\ \\ \\ {$0.3$}}}}
\put(721.0,648.0){\rule[-0.200pt]{0.400pt}{4.818pt}}
\put(120.0,31.0){\rule[-0.200pt]{159.235pt}{0.400pt}}
\put(781.0,31.0){\rule[-0.200pt]{0.400pt}{153.453pt}}
\put(120.0,668.0){\rule[-0.200pt]{159.235pt}{0.400pt}}
\put(12,349){\makebox(0,0){{${m \over {\surd\sigma}}$}}}
\put(450,-53){\makebox(0,0){{$1/N^2_c$}}}
\put(120.0,31.0){\rule[-0.200pt]{0.400pt}{153.453pt}}
\put(621,339){\circle*{12}}
\put(343,321){\circle*{12}}
\put(248,300){\circle*{12}}
\put(242,348){\circle*{12}}
\put(621.0,330.0){\rule[-0.200pt]{0.400pt}{4.577pt}}
\put(611.0,330.0){\rule[-0.200pt]{4.818pt}{0.400pt}}
\put(611.0,349.0){\rule[-0.200pt]{4.818pt}{0.400pt}}
\put(343.0,309.0){\rule[-0.200pt]{0.400pt}{5.782pt}}
\put(333.0,309.0){\rule[-0.200pt]{4.818pt}{0.400pt}}
\put(333.0,333.0){\rule[-0.200pt]{4.818pt}{0.400pt}}
\put(248.0,275.0){\rule[-0.200pt]{0.400pt}{12.045pt}}
\put(238.0,275.0){\rule[-0.200pt]{4.818pt}{0.400pt}}
\put(238.0,325.0){\rule[-0.200pt]{4.818pt}{0.400pt}}
\put(242.0,324.0){\rule[-0.200pt]{0.400pt}{11.563pt}}
\put(232.0,324.0){\rule[-0.200pt]{4.818pt}{0.400pt}}
\put(232.0,372.0){\rule[-0.200pt]{4.818pt}{0.400pt}}
\put(621,479){\circle{12}}
\put(343,440){\circle{12}}
\put(248,431){\circle{12}}
\put(242,463){\circle{12}}
\put(621.0,471.0){\rule[-0.200pt]{0.400pt}{4.095pt}}
\put(611.0,471.0){\rule[-0.200pt]{4.818pt}{0.400pt}}
\put(611.0,488.0){\rule[-0.200pt]{4.818pt}{0.400pt}}
\put(343.0,422.0){\rule[-0.200pt]{0.400pt}{8.913pt}}
\put(333.0,422.0){\rule[-0.200pt]{4.818pt}{0.400pt}}
\put(333.0,459.0){\rule[-0.200pt]{4.818pt}{0.400pt}}
\put(248.0,398.0){\rule[-0.200pt]{0.400pt}{15.658pt}}
\put(238.0,398.0){\rule[-0.200pt]{4.818pt}{0.400pt}}
\put(238.0,463.0){\rule[-0.200pt]{4.818pt}{0.400pt}}
\put(242.0,423.0){\rule[-0.200pt]{0.400pt}{19.513pt}}
\put(232.0,423.0){\rule[-0.200pt]{4.818pt}{0.400pt}}
\put(232.0,504.0){\rule[-0.200pt]{4.818pt}{0.400pt}}
\end{picture}

\end	{center}
\caption{Lightest scalar ($\bullet$) and tensor ($\circ$)
 glueball masses in $D=4$. Continuum values for $N_c=2,3$
 and lattice values ($\beta=10.9$ and $\beta=11.1$)
 for $N_c=4$.}
\label	{fig_plot_glue4}
\end 	{figure}

In Fig.~\ref{fig_plot_glue4} we plot the scalar and tensor glueball
masses, in units of $\surd\sigma$, as a function of $N_c$.
For $N_c=2,3$ we have used the continuum values. For $N_c=4$
the calculations are not precise enough to permit an extrapolation
to the continuum limit and so we simply present the values 
that we obtained at $\beta=10.9$ and 11.1. (We do not use the 
$\beta=10.7$ values since they have large errors and there is the
danger that the scalar mass may be reduced by its proximity to
the critical point at the end of the bulk transition line.)
Although the $N_c=4$ errors are quite large, it certainly seems that
there is little variation with $N_c$ for $N_c\geq 2$ and
any dependence appears to be consistent with being given by a simple
$1/N_c^2$ correction. The fact that these mass ratios appear
to have finite non-zero limits, implies that the large-$N_c$
theory is confining. 

\begin	{figure}[p]
\begin	{center}
\leavevmode
\setlength{\unitlength}{0.240900pt}
\ifx\plotpoint\undefined\newsavebox{\plotpoint}\fi
\sbox{\plotpoint}{\rule[-0.200pt]{0.400pt}{0.400pt}}%
\begin{picture}(825,584)(0,0)
\font\gnuplot=cmr10 at 12pt
\gnuplot
\sbox{\plotpoint}{\rule[-0.200pt]{0.400pt}{0.400pt}}%
\put(120.0,31.0){\rule[-0.200pt]{4.818pt}{0.400pt}}
\put(108,31){\makebox(0,0)[r]{{$0$}}}
\put(761.0,31.0){\rule[-0.200pt]{4.818pt}{0.400pt}}
\put(120.0,177.0){\rule[-0.200pt]{4.818pt}{0.400pt}}
\put(108,177){\makebox(0,0)[r]{{$0.2$}}}
\put(761.0,177.0){\rule[-0.200pt]{4.818pt}{0.400pt}}
\put(120.0,322.0){\rule[-0.200pt]{4.818pt}{0.400pt}}
\put(108,322){\makebox(0,0)[r]{{$0.4$}}}
\put(761.0,322.0){\rule[-0.200pt]{4.818pt}{0.400pt}}
\put(120.0,468.0){\rule[-0.200pt]{4.818pt}{0.400pt}}
\put(108,468){\makebox(0,0)[r]{{$0.6$}}}
\put(761.0,468.0){\rule[-0.200pt]{4.818pt}{0.400pt}}
\put(320.0,31.0){\rule[-0.200pt]{0.400pt}{4.818pt}}
\put(320,19){\makebox(0,0){\shortstack{\\ \\ \\ {$0.1$}}}}
\put(320.0,557.0){\rule[-0.200pt]{0.400pt}{4.818pt}}
\put(521.0,31.0){\rule[-0.200pt]{0.400pt}{4.818pt}}
\put(521,19){\makebox(0,0){\shortstack{\\ \\ \\ {$0.2$}}}}
\put(521.0,557.0){\rule[-0.200pt]{0.400pt}{4.818pt}}
\put(721.0,31.0){\rule[-0.200pt]{0.400pt}{4.818pt}}
\put(721,19){\makebox(0,0){\shortstack{\\ \\ \\ {$0.3$}}}}
\put(721.0,557.0){\rule[-0.200pt]{0.400pt}{4.818pt}}
\put(120.0,31.0){\rule[-0.200pt]{159.235pt}{0.400pt}}
\put(781.0,31.0){\rule[-0.200pt]{0.400pt}{131.531pt}}
\put(120.0,577.0){\rule[-0.200pt]{159.235pt}{0.400pt}}
\put(12,400){\makebox(0,0){{${{\chi_t^{1\over 4}} \over {\surd\sigma}}$}}}
\put(450,-53){\makebox(0,0){{$1/N^2_c$}}}
\put(120.0,31.0){\rule[-0.200pt]{0.400pt}{131.531pt}}
\put(621,373){\circle*{12}}
\put(343,349){\circle*{12}}
\put(245,327){\circle*{12}}
\put(245,254){\circle*{12}}
\put(621.0,359.0){\rule[-0.200pt]{0.400pt}{6.745pt}}
\put(611.0,359.0){\rule[-0.200pt]{4.818pt}{0.400pt}}
\put(611.0,387.0){\rule[-0.200pt]{4.818pt}{0.400pt}}
\put(343.0,331.0){\rule[-0.200pt]{0.400pt}{8.672pt}}
\put(333.0,331.0){\rule[-0.200pt]{4.818pt}{0.400pt}}
\put(333.0,367.0){\rule[-0.200pt]{4.818pt}{0.400pt}}
\put(245.0,306.0){\rule[-0.200pt]{0.400pt}{10.118pt}}
\put(235.0,306.0){\rule[-0.200pt]{4.818pt}{0.400pt}}
\put(235.0,348.0){\rule[-0.200pt]{4.818pt}{0.400pt}}
\put(245.0,221.0){\rule[-0.200pt]{0.400pt}{16.140pt}}
\put(235.0,221.0){\rule[-0.200pt]{4.818pt}{0.400pt}}
\put(235.0,288.0){\rule[-0.200pt]{4.818pt}{0.400pt}}
\end{picture}

\end	{center}
\caption{The topological susceptibility: continuum values
 for $N_c=2,3$ and lattice values ($\beta=10.9$ and $\beta=11.1$) 
 for $N_c=4$.}
%
\label	{fig_plot_top4}
\end 	{figure}
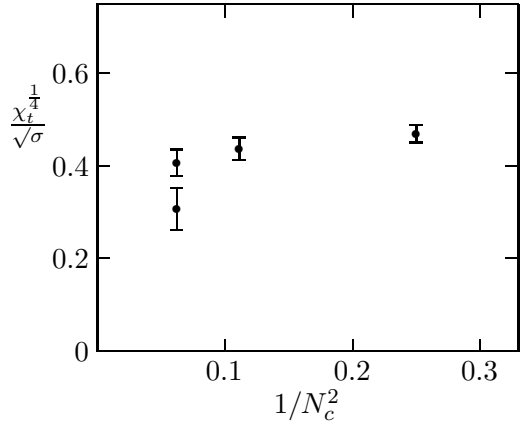

As mentioned earlier we have also calculated the topological
susceptibility. In Fig.~\ref{fig_plot_top4} we plot 
the dimensionless ratio $\chi_t^{1/4}/\surd\sigma$ as a
function of $N_c$. Once again the $N_c$=2 and 3 values
are continuum extrapolations of lattice values
\cite{CMT,Qsu3},
while in the case of $SU(4)$ we simply display the lattice
values obtained at $\beta$=10.9 and 11.1.
We remark that for $SU(4)$ one expects, semiclassically, very 
few small instantons and this is confirmed in our cooling
calculations. This has the advantage that the 
lattice ambiguities that arise when instantons 
are not much larger than $a$
are reduced as compared to $SU(3)$, and dramatically reduced
as compared to $SU(2)$. This implies that the interesting 
large-$N_c$ physics of topology (and the related meson physics)
should be straightforward to study.

\section	{Conclusions.}

\label		{conc}

We have calculated the mass spectra and string tensions of 
gauge theories with $N_c=2,3,4$ in 3 dimensions. We have 
found that there is only a small variation with $N_c$
and this can be accurately described by a modest $O(1/N_c^2)$ 
correction. That is to say, such theories are close to their 
$N_c=\infty$ limit for all values of $N_c \geq 2$. We find that
the large-$N_c$ theory is confining and that $g^2 \propto 1/N_c$
when expressed in physical units. This confirms, in a fully
non-perturbative way,  expectations arrived at from analyses
of Feynman diagrams. It simultaneously provides a unified understanding of 
all our $SU(N_c)$ theories in terms of just the one theory, 
$SU(\infty)$, with modest corrections to it. In practical terms
this means that, from the parameters in our Tables, we know the 
corresponding masses for $all$ values of $N_c$.

Our calculations in 4 dimensions, while quite preliminary, 
suggest that the situation is the same there as in 3 dimensions.

\vskip 0.50in

{\leftline{\large{\bf{Acknowledgements}}}}

We are grateful to PPARC for support under GR/K55752 and for
computing support under GR/J21408 and GR/K95338.

\vfill
\eject

\begin 	{table}[p]
\begin	{center}
\begin	{tabular}
{|c| r@{.}l@{ (}r@{) }|}
\hline
\multicolumn{1}{|c|}{$N_c$} & 
\multicolumn{3}{|c|}{${{\surd\sigma}/{g^2}}$} \\
\hline
 2 & 0&3350 &  15 \\
 3 & 0&5530 &  20 \\
 4 & 0&7564 &  45 \\
\hline
\end	{tabular}
\end	{center}
\caption{The $D=2+1$ $SU(N_c)$ confining string tension.}
\label	{n_string}
\end 	{table}

\begin 	{table}[p]
\begin	{center}
\begin	{tabular}
{|c| r@{.}l@{ (}r@{) }| r@{.}l@{ (}r@{) }|c|}
\hline
\multicolumn{1}{|c|}{$\beta$} &
\multicolumn{3}{|c|}{$R_{\infty}$} & 
\multicolumn{3}{|c|}{$R_{slope}$} & \multicolumn{1}{|c|}{CL$(\%)$} \\
\hline
 $0^{++}$      & 0&805 &  13 & 	-0&06 &  8 & 	  90  \\
 $0^{++\star}$ & 1&245 &  27 & 	-0&41 & 14 & 	  85  \\
 $0^{-+}$      & 1&788 &  88 & 	-0&48 & 56 & 	  40  \\
 $2^{++}$      & 1&333 &  29 & 	-0&08 & 18 & 	  45  \\
 $2^{-+}$      & 1&340 &  40 & 	-0&01 & 24 & 	  12  \\
 $1^{++}$      & 1&946 &  75 & 	-0&59 & 47 & 	  95  \\
 $1^{-+}$      & 1&919 & 115 & 	-0&18 & 75 & 	  30  \\
\hline
\end	{tabular}
\end	{center}
\caption{States with $C=+$ in $D=2+1$ : $R_{\infty} \equiv 
{\displaystyle\lim_{N_c\to\infty}{{m_G}\over{g^2N_c}}}$
and $R_{slope}$ is the coefficient of the $1/N^2_c$ 
correction in eqn(4).}
\label	{ng_plus}
\end 	{table}
\begin 	{table}[p]
\begin	{center}
\begin	{tabular}
{|c| r@{.}l@{ (}r@{) }| r@{.}l@{ (}r@{.}l@{) }|}
\hline
\multicolumn{1}{|c|}{$G$} & 
\multicolumn{3}{|c|}{$R_{\infty}$} & 
\multicolumn{4}{|c|}{$R_{slope}$}\\
\hline
 $0^{--}$      & 1&18 &  6 & 	 0&1 &  0&6 \\
 $0^{--\star}$ & 1&47 & 10 & 	 0&3 &  1&1 \\
 $0^{+-}$      & 1&98 & 28 & 	-0&4 &  2&7 \\
 $2^{--}$      & 1&52 & 14 & 	 0&9 &  1&4 \\
 $2^{+-}$      & 1&58 & 13 & 	-0&4 &  1&3 \\
 $1^{--}$      & 1&85 & 15 & 	-0&3 &  1&5 \\
 $1^{+-}$      & 1&78 & 23 & 	 1&3 &  2&3 \\
\hline
\end	{tabular}
\end	{center}
\caption{As in Table 2 but for states with $C=-$.}
\label	{ng_minus}
\end 	{table}
\begin 	{table}[p]
\begin	{center}
\begin	{tabular}
{|c| r@{.}l@{ (}r@{) }|r@{.}l@{ (}r@{) }|r@{.}l@{ (}r@{) }|}
\hline
\multicolumn{1}{|c|}{$\beta$} &
\multicolumn{3}{|c|}{$a\surd\sigma$} & 
\multicolumn{3}{|c|}{$am_{0^{++}}$} & 
\multicolumn{3}{|c|}{$am_{2^{++}}$} \\
\hline
 10.7  &  0&296 & 14 & 	0&98 & 17 & 1&78 & 34 \\
 10.9  &  0&229 &  7 & 	0&77 &  8 & 1&20 & 10 \\
 11.1  &  0&196 &  7 & 	0&78 &  6 & 1&08 & 10 \\
\hline
\end	{tabular}
\end	{center}
\caption{$SU(4)$ in 4 dimensions; masses calculated at
the values of $\beta$ shown.}
\label	{ngk_4d}
\end 	{table}
%
%
%
%
%

\begin{thebibliography}{99}

\bibitem{tHo}
G. 't Hooft,
Nucl. Phys. B72 (1974) 461.

\bibitem{Wit}
E. Witten,
Nucl. Phys. B160 (1979) 57.

\bibitem{Cole}
S. Coleman,
1979 Erice Lectures.

\bibitem{EK}
T. Eguchi and H. Kawai,
Phys. Rev. Lett. 48 (1982) 1063.

\bibitem{Das}
S.R. Das,
Rev. Mod. Phys. 59(1987)235.

\bibitem{MT2G}
M. Teper,
Phys. Lett. B289 (1992) 115.

\bibitem{MT3G}
M. Teper,
in preparation.

\bibitem{patisg}
N. Isgur and J. Paton,
Phys. Rev. D31 (1985) 2910.

\bibitem{tepmor}
T. Moretto and M. Teper,
heplat-9312035.

\bibitem{Dalley}
F. Antonuccio and S. Dalley,
Nucl. Phys. B461 (1996) 275.

\bibitem{MT2K}
M. Teper,
Phys. Lett. B311 (1993) 223.

\bibitem{MTLAT96}
M. Teper,
to appear in the Proceedings of Lattice '96.

\bibitem{CMT}
C. Michael and M. Teper,
Nucl. Phys. B305 (1988) 453; B314 (1989) 347.

\bibitem{tcsu2}
M. Teper,
Phys. Lett. B313 (1993) 417.

\bibitem{Qsu3}
M. Teper,
Phys. Lett. B202 (1988) 553.

\bibitem{MFI}
G. Parisi, in {\it High Energy Physics} - 1980(AIP 1981);
G. Lepage and P. Mackenzie, Phys. Rev. D48 (1993) 2250.

\end{thebibliography}
\end{document}